# Screening Current-Induced Field and Field Drift Study in HTS coils using *T-A* homogenous model

Edgar Berrospe-Juarez[1], Frederic Trillaud[2], Víctor M R Zermeño[3], and Francesco Grilli[4]

[1]Postgraduate School of Engineering, National Autonomous University of Mexico, Mexico

[2]Institute of Engineering, National Autonomous University of Mexico, Mexico

[3]NKT, Germany

[4]Karlsruhe Institute of Technology, Germany

Email: eberrospej@iingen.unam.mx

**Abstract:** The emergence of second generation (2G) high-temperature superconductor (HTS) tapes has favored the development of HTS magnets for their applications in areas such as NMR, MRI and high field magnets. The screening current-induced field and the field drift are two major problems hindering the use of HTS tapes in the mentioned areas. Both problems are caused by the screening current, then it is necessary to have a modeling strategy capable to estimate such phenomena. Thus far, the *H* formulation has been the most successful and used approach to model medium-size systems (hundreds of tapes). However, its application to large-scale systems is still impaired by excessive computation times and memory requirements. Homogenization and multi-scaling strategies have been successfully implemented to increase the computational efficiency. In this contribution, we show that using the homogenization technique with the recently developed *T-A* formulation allows reducing the computation time and the amount of memory up to the point that real-time simulations of slow ramping cycles of large-scale systems are possible. The *T-A* homogeneous model also allows systematically investigating the screening current using numerical simulations

Keywords: large-scale superconductor systems, HTS coils, screening current, screening current-induced field, field drift.

## 1. Introduction

The quality of second-generation high-temperature superconductor (HTS) tapes has improved during recent years. Now that HTS tapes are commercially available in sufficient length, there is an increasing interest in the development of devices based on this technology. The applications include power system devices [1], NMR and MRI [2-4], accelerators for medical applications [5], scientific magnets [6-8]. The HTS materials are performing so well that they preserve their high critical current even under the effect of high magnetic fields that exceed the critical fields of low-temperature superconductors. For this reason, they are good candidates to produce magnets above 20 T. Some projects around the world are moving in this direction [7, 9, 10].

In coils wound with HTS tapes, the large aspect ratio (width/thickness) of the tapes favours the presence of screening currents. These currents are induced by the gradual penetration of the radial component of the magnetic field. The magnetic field generated by screening currents affects the quality of the overall magnetic field in terms of field reduction by the shielding current and temporal stability of the central field. These two issues are known as screening current-induced field (SCIF) and central field drift [2, 4, 11].

Some of these applications, like accelerator, NMR and MRI magnets, require a high-accuracy of the magnetic field in a fiducial volume located around the center of the magnet [3, 5]. Therefore, phenomena like the SCIF and the field drift need to be understood and remedied. For such purpose, numerical models are an irreplaceable tool. Finite element models based on the *H* formulation have been successfully used during the last years to model systems made with HTS tapes [12]. The *H* formulation models are limited to the study of medium-size systems, because the study of large-scale systems requires an excessive amount of computational resources. This limitation has favored the emergence of the homogenization [13] and multi-scale methods [14, 15]. A major step towards the reduction of the required computational resources was undergone with the use of the *T-A* formulation [16, 17]. More recently, *T-A* homogenous and *T-A* multi-scale strategies have further reduced the computation time and the amount of memory up to the point that it is possible to achieve real-time simulations of slow ramping cycles of large-scale superconducting systems [18].

In this manuscript, we present the *T-A* homogenous model of an HTS solenoidal coil. This coil model is used to qualitatively investigate the dependence of the SCIF and field drift on the critical current density ($J_c$) and the $n$ power law index. The effect of $J_c$ has already been addressed in the literature [2, 19]. Here, the effects of $n$ are included in the discussion. Indeed, $n$, similarly to $J_c$, depends on quantities such as the magnetic field, the temperature and the stress [20, 21], and, therefore, its variation could have an important impact as well. The *T-A* homogenous model also allows to study the effectiveness of two remedies for the SCIF and field drift: the striation of the tapes [11] and the current sweep reversal [2, 22].

## 2. *T-A* homogenous approach

In this section, a brief description of the *T-A* homogenous approach is presented. Comprehensive descriptions of the *T-A* formulation can be found in [16, 17], while the reference of the *T-A* homogenous approach is given in [18].

The so-called *T-A* formulation is the combination of the *T* and the *A* formulations. The HTS layer of the tapes are considered to be one dimensional (1D) lines. The current vector potential (**T**) is exclusively defined in the HTS 1D layers, and the magnetic vector potential (**A**) is defined over the entire system. The two vector potentials are defined by the following relations,

$$\mathbf{J} = \nabla \times \mathbf{T} \tag{1}$$

$$\mathbf{B} = \nabla \times \mathbf{A} \tag{2}$$

where **B** and **J** are the magnetic flux density and the current density, respectively [12]. The governing equations of the *T* and *A* formulation are,

$$\nabla \times \rho \nabla \times \mathbf{T} = -\frac{\partial \mathbf{B}}{\partial t}, \tag{3}$$

$$\nabla \times \nabla \times \mathbf{A} = \mu \mathbf{J}, \tag{4}$$

where $\rho$ is the resistivity and $\mu$ is the magnetic permeability.

The study of solenoidal coils can be addressed by means of a 2D axisymmetric model, see figure 1, where **T** and **A** have only one component. Consequently, equations (3) and (4) are simplified to,

$$\nabla^2 A_\varphi = \mu J_\varphi \tag{5}$$

$$\frac{\partial}{\partial z}\left(\rho_{HTS}\frac{\partial T_r}{\partial z}\right) = \frac{\partial B_r}{\partial t} \tag{6}$$

where $\rho_{HTS}$ is the resistivity of the HTS material.
The transport current flowing through the tapes is imposed by the Dirichlet boundary condition on $T_r$, that must satisfy the following relation,

$$I = (T_1 - T_2)\delta \tag{7}$$

where $I$ is the transport current, $T_1$ and $T_2$ are the $T_r$ values at the edges of the superconducting line, and $\delta$ is the thickness of the HTS layer. The latter being infinitely thin, equation (7) imposes a *sheet current density* $I/\delta$ in practice.

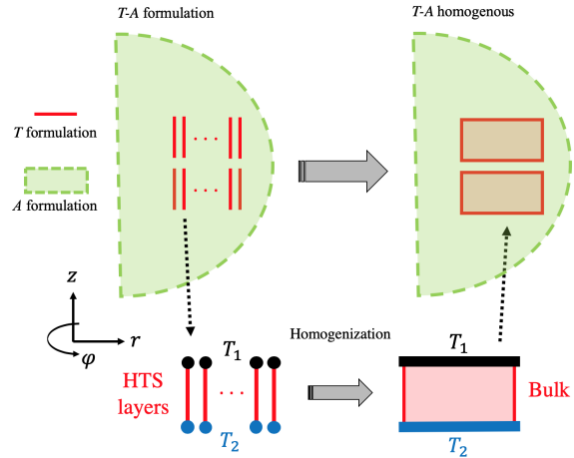

**Figure 1.** 2D axisymmetric representation of a solenoidal coil where the HTS layers are considered to be 1D lines. The *T-A* formulation combines the *T* and *A* formulations, T is defined just along the HTS lines. The homogenization process transforms the stacks of lines into homogeneous bulks.

The homogeneous model assumes that a stack of HTS tapes can be represented by a homogeneous bulk, such process is depicted in figure 1. The bulk can be understood as the limit case of a stack where the spaces between adjacent HTS layers disappear. Analogous to the non-homogenous case, **T** is now exclusively defined inside the bulks. Equation (6) is valid again, because the influence of the magnetic field component parallel to the surface of the tapes on the computation of **T** is neglected. Such component, however, influences the local value of $J_c$ (see equation (9)). The HTS bulks are considered to be surrounded with a medium of zero electrical conductivity, as proposed in [18].

## 3. Case study

The case study used in this manuscript is a solenoidal coil made of 10 HTS pancakes, already described in [15]. Each pancake coil have 80 turns and each turn has same dimensions. Tables 1 and 2 contain the relevant parameters of the coil and of the HTS tape.

**Table 1.** Coil parameters.

| Parameter | Value |
|---|---|
| Pancakes | 10 |
| Turns/pancake | 80 |
| Inner radius | 20 mm |
| Outer radius | 40 mm |
| Height | 44.5 mm |

**Table 2.** HTS tape parameters.

| Parameter | Value |
|---|---|
| Width | 4 mm |
| HTS layer thickness | 1 μm |
| $E_c$ | 1e-4 Vm$^{-1}$ |
| $n$ | 25 |
| $J_{c0}$ | 4.5e10 Am$^{-2}$ |
| $B_0$ | 0.03 T |
| $k$ | 0.2 |
| $\alpha$ | 0.6 |

The electrical resistivity of the HTS material is modeled with the so-called $E$-$J$ power law,

$$\rho_{HTS} = \frac{E_c}{J_c(\mathbf{B})} \left| \frac{\mathbf{J}}{J_c(\mathbf{B})} \right|^{n-1}. \quad (8)$$

It is considered that the critical current density ($J_c$) has an anisotropic dependence on the magnetic field, described by the following elliptic relation,

$$J_C(\mathbf{B}) = \frac{J_{c0}}{\left(1 + \frac{\sqrt{k^2 B_z^2 + B_r^2}}{B_0}\right)^\alpha}, \quad (9)$$

where $B_z$ and $B_r$ are the magnetic flux density components. The values of the parameters of (8) and (9) are listed in Tables 1 and 2. Since there are no magnetic materials present in the system, its permeability is the one of the vacuum $\mu_0$.

The pancakes, once converted into bulks, is meshed with 60 equally-spaced rectangular elements, along the tape's width. The number of elements along the bulk's thickness is 11, the distribution of these elements is graded with increasing number of elements at the extremities of the bulk. The symmetries of the coil allow employing adequate boundary conditions to model just the right upper quadrant of the full coil. The model was implemented in COMSOL Multiphysics 5.4, with the help of the PDE and AC/DC modules.

## 4. Screening current-induced field

The screening current is induced by the gradual penetration of the magnetic field into the tapes. The magnetic field penetrates from the extremities of the tapes producing current fronts. The screening current affects the magnetic field distribution all over the coil and particularly attenuates the central magnetic field. This attenuation is referred to as screening current-induced field (SCIF) [2], and it is quantified by the following relation,

$$B_{SC} = B_{sim} - B_n \qquad (10)$$

where $B_{sim}$ is the central magnetic flux density estimated by means of the model, here by the *T-A* formulation homogeneous model described above, and $B_n$ is the central nominal magnetic flux density produced by a uniform distribution of current density. For this case study the central point has the coordinates (0, 0).

When $B_{SC}$ is plotted as a function of $B_n$, a hysteresis loop is formed. The simulations in this section consider that the transport current is given by the figure 2. The simulations are extended to one and a half cycles (6 h), this lapse is enough to capture the full hysteretic behavior of the SCIF. The amplitude of the transport current is 26.5 A. This charge cycle represents the real conditions of operations of high-field magnets [9].

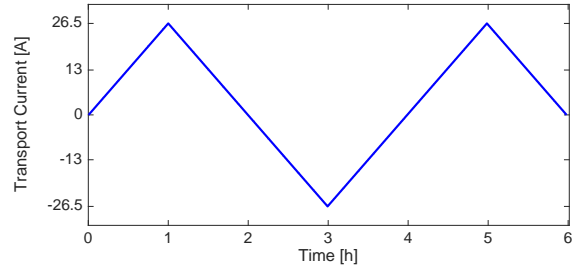

**Figure 2.** Transport current. Triangular shape with a 4 h period, and 26.5 A amplitude.

The magnitude of the magnetic flux density produced when the current density is uniform, at the first current peak ($t = 1$ h), is shown in figure 3 a). The magnetic flux density estimated with the *T-A* homogeneous model, i.e. including the screening current effects, at $t = 1$ h, is shown in figure 3 b). For these conditions $B_n = 0.36$ T and $B_{sim} = 0.345$ T, then $B_{SC} = 0.015$ T ( 4.1% ). Even if it is not easy to appreciate the attenuation of the central field, figure 3 clearly shows how the screening current distorts the field all over the coil. The SCIF is known to depend on the coil's shape and values as large as 18% have been reported in [23].

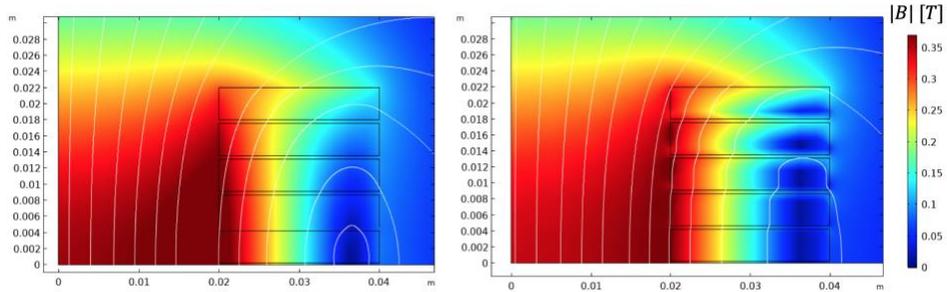

**Figure 3.** a) Magnetic flux density magnitude at peak current produced by a uniform $J$. b) Magnetic flux density magnitude estimated with the *T-A* homogeneous model in the base conditions $\{J_c = J_b, n = 25\}$. The black lines show the position of the pancakes. The white stream lines show the field distortions produced by the screening current.

The effects of $J_c$ and $n$ in the form and size of the SCIF hysteresis loop are studied hereinafter. The $J_{c0}$ and $n$ provided in Table 2 are considered to be the reference values for the rest of the study. Two batches of simulations are performed modifying the base conditions. The first batch of simulations modifies the critical current density by a factor at a constant $n$ whereas the second batch modifies $n$ at a constant $J_{c0}$.

The first batch considers $n = 25$ and $J_{c0n} = [1.5, 1.0, 0.5, 0.25] * J_{c0}$; the critical current density $J_c$ remains as defined in (9) replacing $J_{c0}$ by $J_{c0n}$. Figure 4 shows the normalized current density ($J_n = J/J_c$) at $t = 1$ h for the four new $J_{c0}$ values. Figure 4 b) corresponds to the field shown in figure 3 b). It is possible to observe how $J_c$ influences the current density distribution, when $J_{c0n} = 1.5 * J_{c0}$, the lower pancakes have non-penetrated zones, while when $J_{c0n} = 0.25 * J_{c0}$ the current density is almost uniform, but reaching fully saturated values for the upper pancakes.

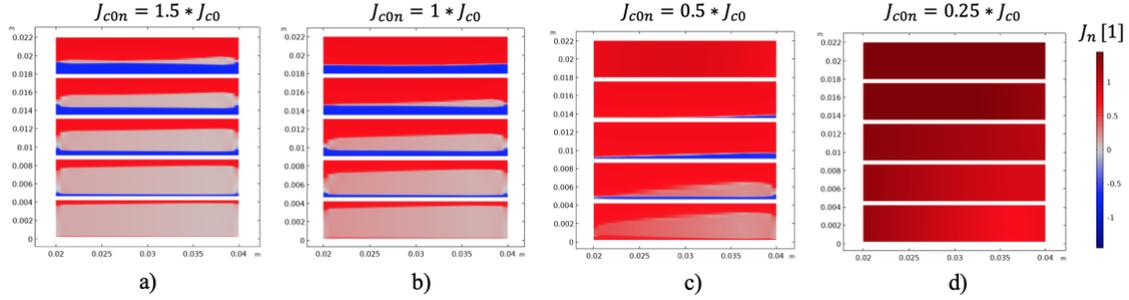

**Figure 4.** Normalized current density ($J_n$) at peak current, for the cases $n = 25$, $J_{c0n} = [1.5, 1.0, 0.5, 0.25] * J_{c0}$. The screening current is clearly observed when $J_{c0n} = 1.5 * J_{c0}$, while an almost uniform $J$ is observed when $J_{c0n} = 0.25 * J_{c0}$.

The SCIF hysteresis loops corresponding to the first simulation batch are presented in figure 5 a). The evolution of the current density can explain the hysteresis loops. At the beginning of the cycle, the current gradually penetrates the tapes incrementing the $B_{SC}$ magnitude. Once the tapes are fully penetrated, $B_{SC}$ tends to saturate. When the magnet is discharged, new current fronts penetrate the tapes and the $B_{SC}$ magnitude decreases, but the declining rate is so that a remnant magnetic field remains at the moment of the zero-crossing point of the cycle. The maximum values for $B_{SC}$ as well as the $B_{SC}$ values at the current peak ($max(B_n)$) are presented in figure 5 b). As it could have already been anticipated from figure 4, figure 5 shows that lower $J_c$ produce lower screening currents and lower $B_{SC}$.

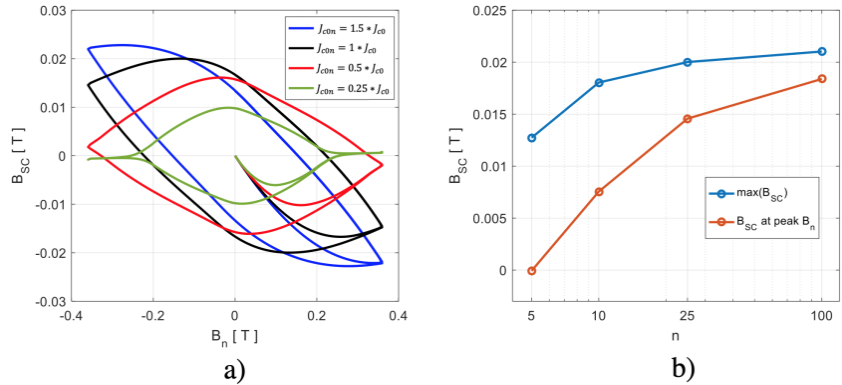

**Figure 5.** a) SCIF hysteresis loop for the cases $n = 25$, $J_{c0n} = [1.5, 1.0, 0.5, 0.25] * J_{c0}$. b) Maximum values of the cases in the left.

The second batch of simulations considers $J_{c0} = 4.5 \times 10^{10} \text{Am}^{-2}$ and $n = [100, 25, 10, 5]$, including thereby the reference case. The SCIF hysteresis loops corresponding to this second batch are presented in figure 6 a). The corresponding maximum values are presented in figure 6 b).

It can be seen that, compared to the impact of the $J_{c0}$, the $n$ has a tendency to round the extremities of the hysteresis loops when the magnetic field changes direction. Similarly to the previous batch, as the $n$ is decreased, the current distribution tends to be more uniform and the hysteresis loops flattens. When $n$ and $J_{c0}$ are decreased, the currents are induced in a larger portion of the HTS layers and the pancakes can saturate at a lower transport current.

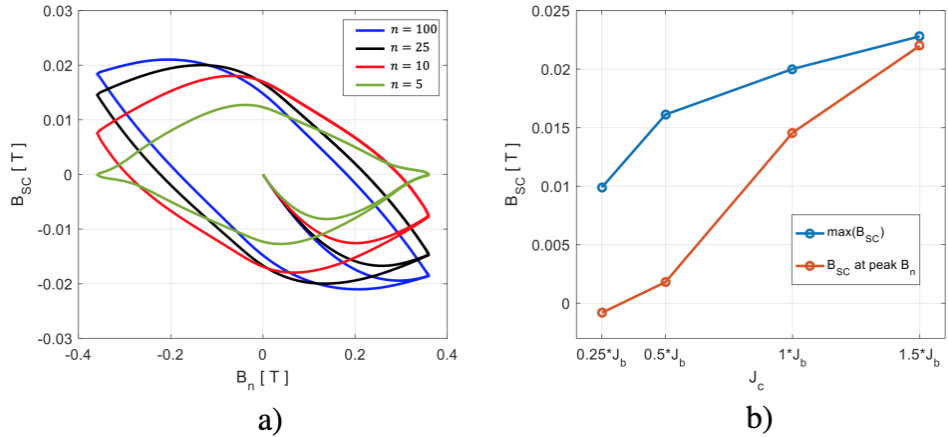

**Figure 6.** a) SCIF hysteresis loop for the cases $J_{c0} = 4.5 \times 10^{10} \text{A.m2}$, $n = [100, 25, 10, 5]$. b) Maximum values of the cases in the left.

## 5. Field drift

Once the coil is charged and the transport current is fixed, the flux creep inside the HTS material causes the relaxation of the screening current and the SCIF, which in turn produces a drift in the central field. In this section, the effect of $J_c$ and $n$ on central field drift is investigated. As in the previous section, the two batches of simulations are considered anew. However, the transport current is a single ramp that reaches a maximum value equal to 26.5 A at $t = 1$ h, followed by a plateau that extends for 30 h to observe the field drift.

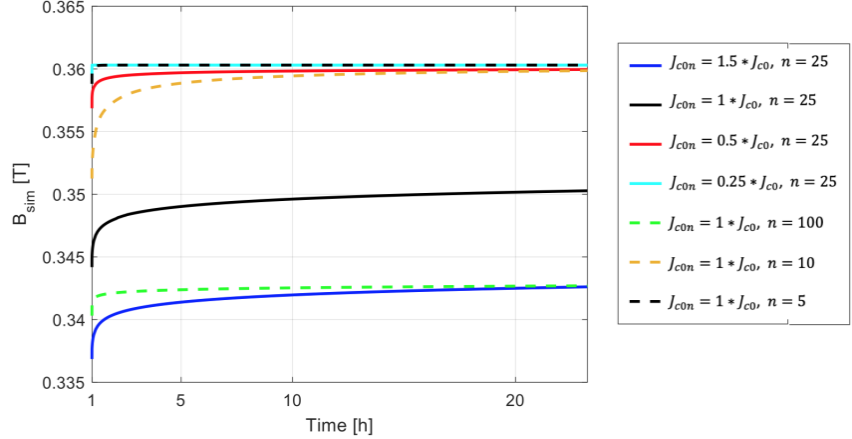

**Figure 7.** Central field drift for the different cases, the plot starts at $t = 1$ h once the maximum current is reached.

Figure 7 shows the estimated $B_{sim}$ for the tested conditions. The *x*-axis has a logarithmic scale which starts at $t = 1$ h and extends to 30 h. The larger drifts occur during the first hours, and a linear growth with logarithmic time is observed during the last hours. The case $\{J_{c0}, n = 10\}$ (orange dashed line) exhibits the larger drift, from 0.352 T to 0.36 T ( 2.2% ). The case $\{1.5 * J_{c0}, n = 25\}$ and $\{J_{c0}, n = 25\}$ (blue and black solid lines) have the same 1.5% drift. The cases represented by the red solid, cyan solid and black dashed lines have a small drift because they are close to the nominal $B_n = 0.36\ T$, and as can be seen in figure 4 d) the $J$ is almost uniform. The case $\{J_{c0}, n = 100\}$ green dashed line has a small 0.4% drift, even when it is away from $B_n$. The conclusion drawn from these simulations is that the size of the drift is a function of the $n$ index, the smaller $n$, the larger the drift. The case with $n = 100$ is closer to the critical state where no flux creep is modeled.

## 6. Striation and current sweep reversal

There are some remedies for the SCIF and the field drift. For example, the SCIF can be reduced by applying an external AC magnetic field forcing the "vortex shaking effect" [24]. A method to reduce the field drift is by lowering the temperature once the operating current is reached, then the critical current is increased and the flux creep diminishes [2]. In this section, two of the common remedies are explored for the SCIF and the field drift problems: i) the striation of the HTS material, and ii) the current sweep. The striation process is the subdivision of the tapes into narrow strips, this process is known to reduce the SCIF [3]. The current sweep reversal is the overshooting of the transport current, this process reduces the field drift [2, 22]

In order to implement the striation in the simulations, the homogeneous bulks, representing the pancakes, are subdivided into narrow pancakes. The boundary conditions for $T$ are modified to force that each new strip forming the narrow pancakes transport a fraction of the original transport current. These conditions correspond to having the strips perfectly uncoupled. The *T-A* homogenous model for the striation case was simulated considering the triangular shape transport current and the reference values used in section 4 and found in Table 2. Figure 8 shows the normalized current density ($J_n$) at $t = 1$ h for the tapes with no striation, taken as a reference case, and for the cases of tapes subdivided into 2, 4 and 8 strips.

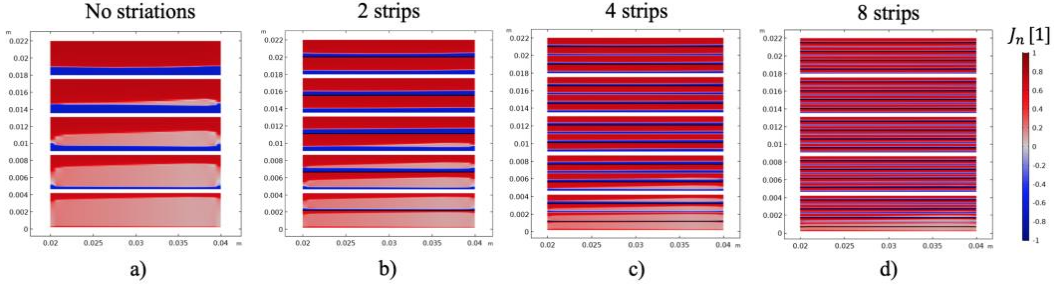

**Figure 8.** Normalized current density ($J_n$) at peak current, for the cases with and without striations.

The striation process causes the formation of current fronts in each strip, and the penetration of the magnetic field from the extremities of the narrow strips produce the overall effect of the reduction of the SCIF. The SCIF hysteresis loops corresponding to the cases with striations are presented in figure 9 a). The maximum values for $B_{SC}$ are presented in figure 9 b).

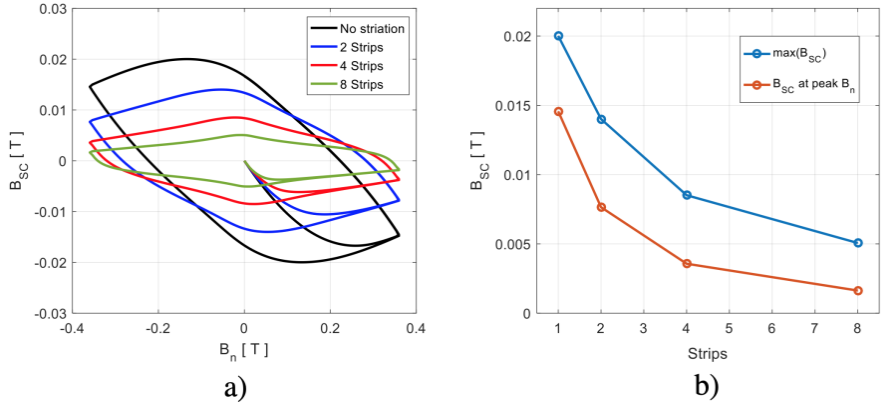

**Figure 9.** a) SCIF hysteresis loop for the cases with striations. b) Maximum values of the cases in the left.

The ramp transport current used in section 5 is modified to implemented the current sweep reversal, the ramp rate is increased, current is overshot and subsequently reduced to reach 26.5 A at $t = 1$ h. Three simulations were performed, one with 5%, 10% and 20% overshoots. These ramps are shown in figure. The reduction of the current after the overshoot produces the appearance of new current fronts at the extremities of the tapes, thus the flux creep and the field drift are reduced.

Figure 10 shows the estimated $B_{sim}$ for the two overshot ramps and the ramp without overshoot. The case with 5% overshoot reduces the drift just during the first minutes. The case with 10% overshoot successfully "eliminates" the drift during the first hours, but around $t = 5.5$ h the drift starts again due to a relaxation of the screening current according to the proposed model. Originally the reappearance of the drift was the motivation to test the 20% overshoot case. It is possible to appreciate that the current sweep reversal with 20% overshoot successfully eliminates the field drift, but during the first minutes a small negative field drift can be appreciated. These results demonstrate that the 1% overshoot reported in [2, 25] is insufficient for our proposed coil and the choice of the overshoot should be determined on a coil basis to address effectively the temporal drift of the central magnetic field. From the lines in figure 10, we can also conclude that the current sweep reversal is an appropriate method to reduce the SCIF.

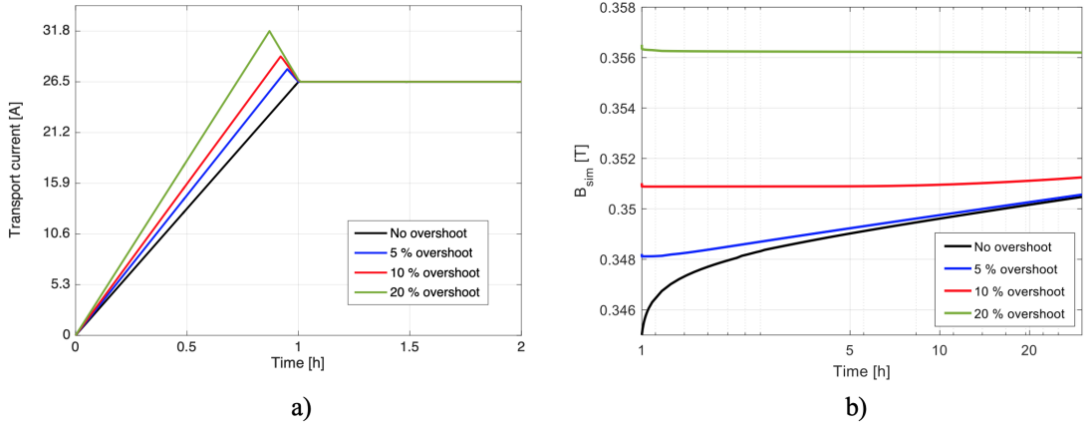

**Figure 10.** a) Ramps with overshoot. b) Estimated central field for the ramps with and without overshoot. The field drift is partially eliminated with 5% and 10% overshoot, and successfully "eliminated" with a 20% overshoot.

The computer used to run the simulations is a desktop computer (6 cores, Intel (R) Xeon(R) ES-2630, 2.2 GHz, 64 GB RAM). The time steps taken by the solver were adjusted to optimize the computation time, this is important because the simulated lapse in the simulations conducted to study the field drift is 3 h, but the time steps can be enlarged when the current is constant. The average computation time required to complete the simulations is 1 h, for both triangular and ramp transport currents. Both simulated lapses, 6 h and 30 h, are larger than the 1 h computation time, therefore, the *T-A* homogeneous models, in the following case studies, could have been performed in real-time during the ramping of the magnet.

## 7. Conclusions

The penetration of the magnetic field into the HTS tapes causes screening currents. The coils wound with HTS tapes face two major problems caused by the screening current, the SCIF and the field drift. These problems, amongst others, must be understood and solved to pave the way for the use of HTS coils in applications like NMR, MRI, high field magnets, etc.

The effects of $J_c$ and $n$ in both problems (SCIF and field drift) were studied by means of several numerical simulations. It was demonstrated that higher $J_c$ values cause higher screening current, while higher $n$ values hinder the relaxation of such screening current. The main conclusion is that the reduction of both quantities $J_c$ and $n$ causes the attenuation of the mentioned problems but they go against the use of high performant HTS tape technologies for which greater critical currents at larger $n$ are sought to the generation of higher magnetic field. Fortunately, the striation of the HTS tapes and the current sweep reversal are effective methods to mitigate the SCIF and the field drift. However, simulations showed that the field drift could reappear for current sweep reversal with 5 and 10% overshoot. This effect could be avoided if the overshoot is increased to 20%. It is likely that the magnitude of the overshoot will depend on the coil configuration as well as the ramping cycle and studies should be performed for each magnet for different operating cycles. The efficiency of the *T-A* homogenous models was exploited in this paper to run dozens of simulations allowing studying the behavior of the magnet during its operation. It should be mentioned that the use of other modeling strategies such as the ones based on the *H* formulation would have required computation times in the order of months to obtain the same conclusion.